\documentclass[12pt]{iopart}
\usepackage{iopams}
\usepackage{epsfig}

\begin{document}

\title[Dabirian {\it et al.}: Propagation of Light in Photonic Crystal Fibre Devices]{Propagation of Light in Photonic Crystal Fibre Devices}

\author{Ali Dabirian\dag\footnote[3]{Corresponding author: Ali Dabirian (ali\_dabirian@yahoo.com) is currently with Patsa
Company, P.O.Box. 14115-337, Tehran, Iran}, Mahmood Akbari\dag\
and Niels Asger Mortensen\ddag }

\address{\dag\ Department of Electrical Engineering, Sharif University of Technology,  Tehran, Iran}

\address{\ddag\ NanoDTU, MIC -- Department of Micro and Nanotechnology, Technical University of Denmark, DK-2800 Kongens Lyngby, Denmark}

\begin{abstract}
We describe a semi-analytical approach for three-dimensional
analysis of photonic crystal fibre devices. The approach relies on
modal transmission-line theory. We offer two examples illustrating
the utilization of this approach in photonic crystal fibres: the
verification of the coupling action in a photonic crystal fibre
coupler and the modal reflectivity in a photonic crystal fibre
distributed Bragg reflector.
\end{abstract}

\submitto{\JOA}

\maketitle

\section{Introduction}
Photonic crystal fibres (PCFs), whose cladding is composed of a
two-dimensional (2D) photonic
crystal~\cite{Yablonovitch:1987,John:1987}, may confine and guide
light through either a photonic bandgap
effect~\cite{Knight:1998,cregan:1999} or by an effective high
refractive index guiding mechanism~\cite{Knight:1996,Birks:1997}.
Both classes of fibres have been the subject of numerous research
and for a review we refer to Refs.~\cite{Russell:2003,Knight:2003}
and references therein. The latter class of PCFs has attractive
features such as, broad-band single-mode (SM)
operation~\cite{Nielsen:2004}, possibilities for dispersion
engineering~\cite{Birks:1999,Knight:2000}, and tailorable mode
area~\cite{Mortensen:2002}. The 2D photonic crystal of the
cladding not only provides more design freedom on engineering
basic properties of the fibre, but also broadens the potential
application of PCFs by the freedom to remove more air holes from
the cladding~\cite{Mangan:2000,Saitoh:2003,Saitoh:2004} or by
introducing additional materials to the air
holes~\cite{Eggleton:2001,Larsen:2003,Larsen:2003a,Du:2004,Alkeskjold:2004},
both of which facilitate novel device operations based on PCFs. In
the present work, the "PCF device" term is restricted to the ones
obtained by removing some holes from the cladding such as, PCF
couplers~\cite{Mangan:2000,Saitoh:2003}, PCF polarization beam
splitters~\cite{Saitoh:2004}, or PCF distributed Bragg
reflectors~\cite{Limpert:2003}.

In the present paper, we describe an approach, which relies on
modal transmission-line theory (MTLT), for three-dimensional (3D)
investigations of the propagation of an optical beam launched into
a PCF device. According to our knowledge only the finite element
beam propagation method (FE-BPM)~\cite{Saitoh:2001} has been
adopted and utilized for doing such simulations. The FE-BPM is
numerically robust, versatile, and applicable to a wide variety of
structures. Unfortunately, this is often achieved at the expense
of long computational times and large memory requirements, both of
which can become critical issues especially when structures with
large dimensions are
considered or when used within an iterative design environment. \\
MTLT, which has been developed for modelling multi-layered
periodic
media~\cite{Peng:1975,Tamir:1996,Akbari:1999,Akbari:2000,Lin:2002},
has been used for analysis of distributed feedback (DFB)
lasers~\cite{Akbari:1999,Akbari:2000}, quantum well infrared
photodetectors (QWIP)~\cite{Yan:1999}, holographic power
splitter/combiners~\cite{Shahabadi:1997} and grating assisted
integrated optics devices~\cite{Zhang:1996}. Recently, it has also
successfully been applied in a study of radiation fields from
end-facet PCFs~\cite{Dabirian:2005b}. MTLT relies on a plane-wave
expansion of electromagnetic fields in the periodic media.
Interpreting the plane waves as transmission-lines provides a
systematic framework for study of wave propagation in
multi-layered periodic media. Besides that, one can exploit all
the concepts and methods of transmission-line
theory~\cite{Faria:1993} and electrical network
theory~\cite{Desoer:1969} for the study of wave propagation. MTLT
has recently been developed for modal analysis of arbitrary shape
optical waveguides~\cite{Dabirian:2005a}. Here, we add a novel
approach to this theory and utilize it for a three-dimensional
study of propagation of light in photonic crystal fibre devices.

The remaining part of the paper is organized as follows. In
Section 2, we give a brief account of MTLT and describe the
approach we use. In Section 3 we investigate examples that
illustrate the utilization of this approach in the modelling of
PCF devices. Finally, conclusions are given in Section 4.
\section{Formalism}
\begin{figure}
\begin{center}
\epsfig{file=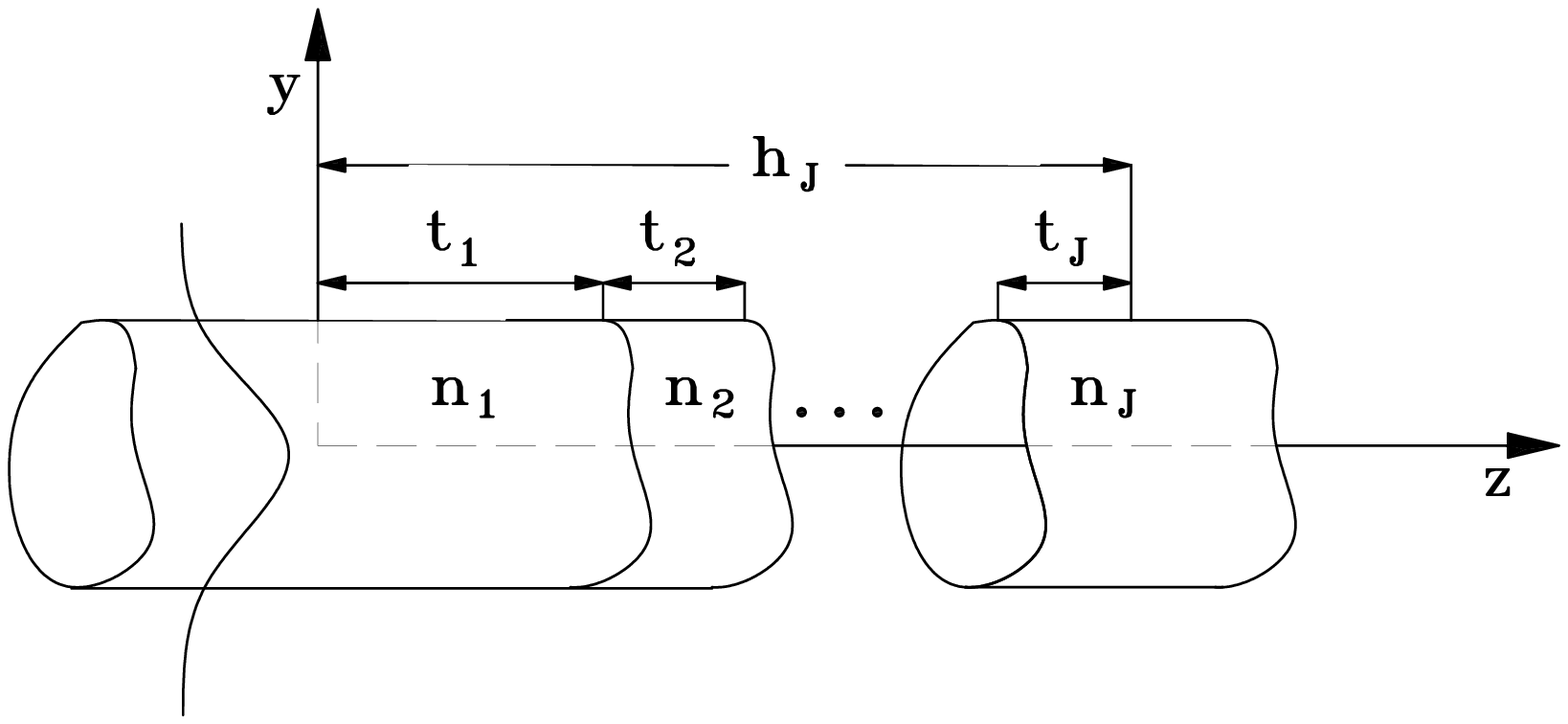,width=\columnwidth}
\end{center}
\caption{General case of a three-dimensional multi-layered
structure.} \label{fig1}
\end{figure}

The typical PCF device that we have in mind is composed of $J$
layers with different relative permittivity functions
$\epsilon_{rj}(x,y),~j=1,2,\cdots,J$, which is illustrated in
 \fref{fig1}. We define $z_j$ as a convenient local
coordinate obeying $0 \leq z_j=z-h_{j-1}\leq t_j$ and we consider
wave propagation along the longitudinal direction of the
structure, i.e. $z$-axis. Throughout this paper we consider
non-magnetic materials with relative permeability $\mu_r=1$ and
 all electromagnetic fields have a harmonic temporal dependence, $\exp(i\omega t)$.
PCF devices with the typical shape shown in \fref{fig1}, usually
confine light within the fibre core or cores. However, some
application such as long-period fibre Bragg gratings do not share
this feature so in the present study we exclude these
applications. For the applications with spatially localized modes
we use a super-cell approach and repeat the structure in the
transverse $xy$-plane, along $x$ and $y$ directions with
periodicities of $T_x$ and $T_y$, respectively. It is assumed that
the periodically repeated devices are separated by a sufficient
amount of background region, here microstructured cladding, that
their electromagnetic fields do not affect each other
significantly.

We want to study a PCF device when it is illuminated with an
incident field ${\vec E}_{inc}({\vec r})$ propagating in layer 1
of the structure shown in \fref{fig1}. This incident field is a
solution to the source-free Maxwell equations in the PCF with the
refractive index profile of layer 1. In this section we briefly
address MTLT and modal analysis of optical waveguides using this
theory. Subsequently we describe an approach, based on MTLT, for
investigating the scattering and propagation of light in PCF
devices. Throughout the paper, vectorial components are denoted by
an arrow placed above them. The bold-style notation with uppercase
and lowercase characters is used to designate matrices and
vectors, respectively.
\subsection{Modal Transmission-Line Theory}
Embody a periodic medium with permittivity variation
$\epsilon_0\epsilon_r(x,y)=\epsilon_0\epsilon_r(x+T_x,y+T_y)$ and
permeability $\mu_0$. The permittivity can then conveniently be
expressed in the form of a two-dimensional Fourier
series~\cite{Akbari:2000}
\begin{equation}\label{g1}
  \epsilon_r(x,y)=\lim_{M,N \longrightarrow \infty}
  \sum_{m=-M}^{M}
  \sum_{n=-N}^{N}
  {\tilde{\epsilon}}_{mn}
  \exp\Big({-im\frac{2\pi}{T_x}x}\Big)
  \exp\Big({-in\frac{2\pi}{T_y}y}\Big)
\end{equation}
where
\begin{equation}\label{g2}
  {\tilde{\epsilon}}_{mn}=\frac{\displaystyle{1}}{\displaystyle{T_xT_y}}
  \int_{0}^{T_x}
  \int_{0}^{T_y}
  \epsilon_r(x,y)
  \exp\Big({im\frac{2\pi}{T_x}x}\Big)
  \exp\Big({in\frac{2\pi}{T_y}y}\Big)dydx~.
\end{equation}
The electromagnetic fields must of course reflect the periodicity
of $\epsilon_r(x,y)$ and according to the Floquet--Bloch theorem
the fields in the doubly periodic medium are pseudo-periodic
functions~\cite{Akbari:2000}
\begin{eqnarray}\label{g3}
\fl{ {{\vec A}}({\vec r})=\lim_{M,N \longrightarrow \infty}
\sum_{m=-M}^{M}  \sum_{n=-N}^{N}  {{\vec a}_{mn}(z)}}
\exp\Big[{-i\Big(K_{x0}+m\frac{2\pi}{T_x}}\Big)x\Big]
\exp\Big[{-i\Big(K_{y0}+n\frac{2\pi}{T_y}}\Big)y\Big]
\end{eqnarray}
where, ${\vec {K}_0}=K_{x0}\hat{x}+K_{y0}\hat{y}$ is the Bloch
wave-vector and ${\vec A}$ can be any of the electromagnetic
fields ${\vec E},~{\vec H}$, or ${\vec D}$. In order to facilitate
calculations in matrix form, we introduce ${\bf \vec e},~{\bf \vec
h}$, and ${\bf \vec d}$ vectors whose elements are ${\vec
e}_{mn},~{\vec h}_{mn}$, and ${\vec d}_{mn}$, respectively. The
dimension of each vectorial component of the ${\bf \vec e},~{\bf
\vec h}$, or ${\bf \vec d}$ vectors in Cartesian coordinates (i.e.
${\bf e_x}$, ${\bf e_y}$, ${\bf e_z}$, ${\bf h_x}$, etc.) is
$1\times (2N+1)(2M+1)$. Using these vectors, the constitutive
relation ${\vec D}=\epsilon_0 \epsilon_r{\vec E}$ converts into
${\bf \vec d}=\epsilon_0 {\bf N} {\bf \vec e}$, where ${\bf N}$ is
a square matrix whose elements are $\tilde{\epsilon}_{mn}$ and
they are arranged in ${\bf N}$ in such a way that the equality
${\bf\vec d}=\epsilon_0 {\bf N} {\bf \vec e}$ holds.\\ The
temporal harmonic electromagnetic fields in a dielectric medium
are solutions of the following source-free Maxwell equations
\begin{eqnarray} \label{g4}
   \left\{
   \begin{array}{r r r}
   \nabla \times {\vec E}({\vec r})&
   =& -i\omega\mu_0{\vec H}({\vec r}) \\
   \nabla\times {\vec H}({\vec r})&
   =  & i\omega{\vec D}({\vec r})
    \end{array}
    \right.
\end{eqnarray}
Using \eref{g1}, \eref{g3}, and vectors ${\bf\vec e},~{\bf\vec h}$
and ${\bf\vec d}$ in the source-free Maxwell equations (\ref{g4}),
these equations are transformed into the following system of
differential equations:
\begin{eqnarray} \label{g5} \left\{  \begin{array}{r r r}
\frac{\displaystyle {d {\bf v}}}{\displaystyle{dz}} &=& -i\omega
{\bf L} {\bf i}
\\\\ \frac{\displaystyle{ d{\bf i}}}{\displaystyle {dz}} &=& -i\omega {\bf C} {\bf v}\end{array}
\right.
\end{eqnarray}
or
\begin{eqnarray} \label{g6} \left\{
  \begin{array}{r r r}
\frac{\displaystyle {d^2{\bf v}}}{\displaystyle {dz^2}}  &=&
-\omega^2 {\bf L} {\bf C} {\bf v}\\
\\
\frac{\displaystyle {d^2 {\bf i}}}{\displaystyle {dz^2}} &=&
-\omega^2 {\bf C} {\bf L} {\bf i}
\end{array}
  \right.
\end{eqnarray}
where ${\bf L}$ and ${\bf C}$ are obtained in the
calculations~\cite{Akbari:2000} and
\begin{equation}
{\bf v}=\left[ \begin{array}{c} {\bf e_y} \\ {\bf e_x} \end{array}
\right], ~{\bf i} = \left[ \begin{array}{c} {\bf h_x} \\ -{\bf
h_y}
\end{array} \right]~.
\end{equation}
\Eref{g5} has the well-known form of telegraphist's equations for
a multi-conductor transmission-line~\cite{Faria:1993} and we have
emphasized the analogy by the choice of symbols so that e.g. ${\bf
i}$ and ${\bf v}$ are interpreted as effective currents and
voltages, respectively. Likewise, inductance and capacitance
matrices of the multi-conductor transmission-line are denoted by
${\bf L}$ and ${\bf C}$, respectively. In equation~(\ref{g6}),
$\omega^2 {\bf LC}$ and $\omega^2 {\bf CL}$ are matrices with
non-zero off-diagonal elements. We can formally diagonalize
$\omega^2 {\bf LC}$ and $\omega^2 {\bf CL}$ matrices using
relations $\omega^2 {\bf LC}={\bf P K^2P^{-1}}$ and $\omega^2 {\bf
CL}={\bf QK^2Q^{-1}}$, where $ {\bf K^2}$ is a diagonal matrix.
The diagonal elements of ${\bf K^2}$ are eigenvalues of $\omega^2
{\bf LC}$ or $\omega^2 {\bf CL}$. Here, ${\bf P}$ and ${\bf Q}$
are matrices whose columns are the eigenvectors of their relevant
non-diagonal matrices. Once the ${\bf K^2}$ and ${\bf P}$ have
been determined, the matrix ${\bf Q}$ is also given by $ \omega
{\bf CPK^{-1}}$.
\\ From the above discussion it follows that
\eref{g6} may be transformed into
\begin{eqnarray}
\label{g7} \left\{
  \begin{array}{r r r}
\frac{\displaystyle {d^2 {\bf \widehat{v}}}}{\displaystyle {dz^2}}
&=& -{\bf K^2\widehat{v}}
\\
\\
\frac{\displaystyle {d^2 {\bf \widehat{i}}}}{\displaystyle {dz^2}}
&=& -{\bf K^2\widehat{i}}
\end{array}
  \right.
\end{eqnarray}
where
\begin{equation}
\label{g8}
 {\bf {v}}={\bf P} {\bf \widehat{
v}}
 ~~,~~{\bf i}={\bf Q \widehat{i}}.
\end{equation}
In this new basis the transmission-lines are uncoupled and one
may, in analogy with conductance eigen-channels in quantum
transport~\cite{Brandbyge:1997}, think of these new lines as the
eigen-lines of the transmission-line system. Wave propagation in
the periodic medium is described by ${\bf K^2}$, ${\bf P}$, ${\bf
Q}$, see \eref{g7} and \eref{g8}.
\\
Evidently from MTLT, the $\omega^2{\bf LC}$ describes the
propagation characteristics of longitudinal space harmonics.
Eigenvalues of this matrix specify the square values of
propagation constants of space harmonics. The propagation
constants are obtained from the diagonal matrix ${\bf K^2}$
considering the following condition~\cite{Tamir:1996}:
\begin{equation}
\label{g9} \textrm{Im}(K_k)+\textrm{Re}(K_k)<0.
\end{equation}
Electromagnetic fields of each space harmonic with a specified
propagation constant are determined from its relevant eigenvector.
\subsection{Equivalent Network of Multi-Layered Media}
\begin{figure}
\begin{center}
\epsfig{file=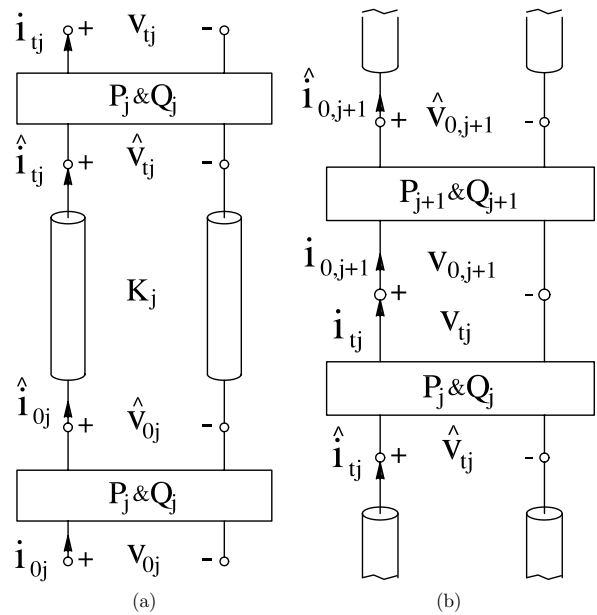,width=0.5\columnwidth}
\end{center}
\caption{Equivalent electrical networks elements (a)
transmission-line unit presenting a single layer (b) General
junction of two transmission-line units at different layer
interface.} \label{fig2}
\end{figure}

\begin{figure}
\begin{center}
\epsfig{file=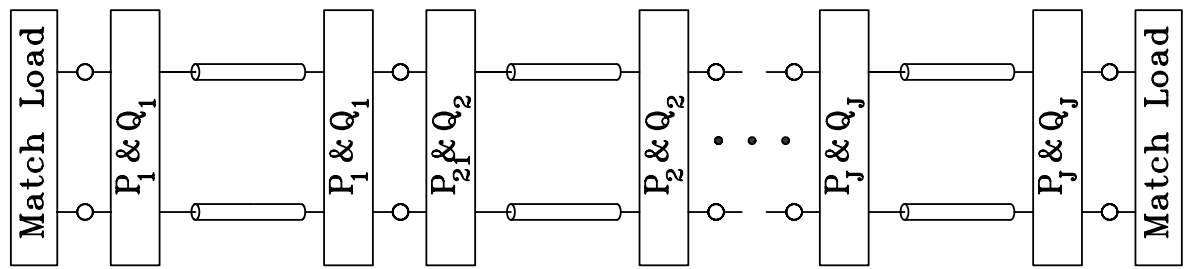,width=\columnwidth}
\end{center}
\caption{Equivalent transmission-line network of the multi-layered
structure shown in Fig.~\ref{fig1}.} \label{fig3}
\end{figure}

Consider the typical structure shown in \fref{fig1}. The modelling
task begins by periodically repeating the device in the transverse
$xy$-plane with sufficiently large periodicities. As discussed
above, wave propagation in each layer of this periodically
repeated structure could be modelled by a transmission-line
network whose behavior is described by \eref{g7}. Schematically,
the equivalent transmission-line network of the {\it j-}th layer
of this structure is depicted in \fref{fig2} (a). In this figure
the box containing $P_j,~Q_j$ represents the consideration in
\eref{g8}.
\\ A concise and effective formulation of voltages and currents of
this transmission-line network can be described by:
\begin{eqnarray}
\label{g10} \left\{
  \begin{array}{r r r}
{\bf \widehat {{v}}_j}& =& {\bf \exp}[-i{\bf K_j}(z_j-t_j) ]{\bf
{\widehat v}_{j,inc}} +{\bf \exp}[i{\bf K_j}(z_j-t_j) ]{\bf
{\widehat v}_{j,r}}\\
\\
{\bf \widehat {{i}}_j}&=& {\bf \exp}[-i{\bf K_j}(z_j-t_j)]{\bf
{\widehat i}_{j,inc}}-{\bf \exp}[i{\bf K_j}(z _j-t_j)]{\bf
{\widehat i}_{j,r}}
\end{array}
  \right.
\end{eqnarray}
where ${\bf{ \widehat v}_{j, inc}},~{\bf {\widehat i}_{j,
inc}},~{\bf {\widehat v}_{j,r}}$, and ${\bf {\widehat i}_{j,r}}$
are vectors for the incident voltage, incident current, reflected
voltage, and reflected current, respectively. The ${\bf
K_j}=\textrm{diag}(K_{j1},K_{j2},\cdots, K_{jk},\cdots)$ is a
diagonal matrix obtained by computing the square root of the ${\bf
K_j^2 }$ matrix. The ${\bf \exp}[-i{\bf K_j} (t_j-z_j)]$ is also a
diagonal matrix with diagonal elements ${\exp[-K_{jk}(t_j-z_j)]}$.
\\ Essential electromagnetic boundary conditions could be simply
satisfied at the interface of two different layers by the
continuity of voltages and currents in transmission-line theory.
At the interface of typical different $l$ and $l+1$ layers,
illustrated in \fref{fig2} (b), the continuity rule is described
by:
\begin{eqnarray}
\label{g11} \left\{
  \begin{array}{r r r}
  {\bf P_j {\widehat v}_{tj}} & = &{\bf P_{j+1}{\widehat v}_{0,j+1}} \\
{\bf Q_j {\widehat i}_{tj}} & = & {\bf Q_{j+1}{\widehat
i}_{0,j+1}}
\end{array}
  \right.
\end{eqnarray}
where ${\bf {\widehat v}_{tj}}$, ${\bf {\widehat v}_{0,j+1}}$,
${\bf {\widehat v}_{tj}}$, and ${\bf {\widehat v}_{0,j+1}}$ have
been defined in \fref{fig2}. On the basis of MTLT the
transmission-line network of the periodically repeated typical PCF
device is illustrated in \fref{fig3}. In the equivalent network of
\fref{fig3} and also in numerical simulation a total height of
$h_J$ is considered. At the beginning ($z=0$) and the end
($z=h_J$) of the structure, the well-known radiation condition of
electromagnetic theory is applied, which is depicted in the
equivalent network by {\it match load}. Here we exploit a primary
feature of radiation condition; i.e. the zero reflection at these
points.
\subsection{ The Approach}
Consider the structure shown in \fref{fig1}. When this structure
is illuminated with an incident electromagnetic field ${\vec
E}_{inc}({\vec r})$, propagating in layer 1 along the positive
direction of $z$-axis, the total field in layer 1, ${\vec
E}_1({\vec r})$, is given by
\begin{equation}\label{g12}
{\vec E}_1({\vec r})={\vec E}_{inc}({\vec r})+{\vec E}_{1,r}({\vec
r})
\end{equation}
where ${\vec E}_{1,r}({\vec r})$ is the reflected field inside
layer 1. The incident field is usually a fibre mode so for
investigating its interaction with other layers we must calculate
it and then calculate ${\vec E}_{1,r}({\vec r})$ and finally
${\vec E}_{1}({\vec r})$. From the known field at layer 1, we
calculate the fields in other layers utilizing equations
\eref{g10} and \eref{g11}.
\\ Since the incident field ${\vec E}_{inc}({\vec r})$ is a guided mode of the waveguide with
refractive index profile of layer 1, we could determine it
utilizing MTLT, exploiting the features of
transmission-lines~\cite{Dabirian:2005a}. This calculation is
achieved by examining the out of plane propagation of a periodic
medium whose refractive index variation is obtained by
periodically repeating the waveguide in transverse plane with
sufficiently large periodicities. Evidently from transmission-line
theory, the matrix $\omega^2 {\bf LC}$ contains the information of
the out of plane propagating waves, called space harmonics in the
field of diffraction grating~\cite{Peng:1975}. Eigenvalues of the
matrix $\omega^2 {\bf LC}$, diagonal elements of ${\bf K^2}$,
specify squared-values of propagation constants of these space
harmonics. Each column of the matrix ${\bf P}$, an eigenvector of
the matrix $\omega^2 {\bf LC}$, describe the electric field
profile of its relevant eigenvalue. Among the space harmonics the
ones whose field profiles are localized within the waveguide
specify guided modes of the waveguide. In index guiding waveguides
this condition is simplified to the refractive index guiding
condition.\\  Through the modal analysis of the fibre with layer 1
refractive index profile we determine ${\bf {\widehat v}_{j,
inc}}(z=0)$. Afterwards, for complete determination of field in
the first layer, calculation of ${\bf {\widehat v}_{j,r}}(z=0)$
and ${\bf {\widehat i}_{j,r}}(z=0)$ is also required. These values
are obtained by the following relations:
\begin{eqnarray}
\label{g13} \left\{\begin{array}{r r r} {\bf {\widehat
v}_{j,r}}(z=0) &= &{\bf R^u_{01}}{\bf {\widehat v}_{j,inc}}(z=0)\\
{\bf {\widehat i}_{j,r}}(z=0)&=&{\bf R^u_{01}}{\bf {\widehat
i}_{j,inc}}(z=0)
\end{array}
  \right.
\end{eqnarray}
where ${\bf R^u_{01}}$ is the upward reflectance matrix at $z=0$.
Generally we define ${\bf R^u_{zj}}$ as the reflectance matrix of
a propagating wave along the positive direction of z-axis at the
local geometry $z_j$; for instance ${\bf R^u_{01}}$ is the upward
reflectance matrix at $z_j=0$ for $j=1$. The variation of ${\bf
R^u_{zj}}$ along $z$ is treated by the following
relation~\cite{Slang:2001}
\begin{eqnarray} \label{g14}
\left\{\begin{array}{r r r}{\bf R_{tj}}& =& F_{rj}({\bf R_{0,j+1}})\\
{\bf R_{oj}}&=&\exp({\bf K_j} x) {\bf R_{tj}}\exp({\bf K_j} x)
\end{array}
  \right.
\end{eqnarray}
where $F_{rj}({\bf R_{0,j+1}})$ is composed of the following set
of equations: \begin{eqnarray}\label{g15} \left\{\begin{array}{r r
r} {\bf R_{tj}}&=&({\bf Z_{tj}-I})({\bf Z_{tj}+I})^{-1} \\ {\bf
Z_{tj}}&=&{\bf P_j^{-1}P_{j+1}Z_{0,j+1}Q^{-1}_{j+1}Q_j} \\
 {\bf Z_{0,j+1}}&=&{\bf (I+R_{0,j+1})(I-R_{0,j+1})^{-1}}
\end{array}
  \right.
\end{eqnarray}
Computation of ${\bf R^u_{01}}$ is started from the topmost layer,
where the reflectance is zero. Considering \eref{g14} and
\eref{g15} at each layer and interface of layers, the ${\bf
R^u_{01}}$ would be calculated. From the known ${\bf R^u_{01}}$,
the ${\bf {\widehat v_{1,r}}}$ and ${\bf {\widehat i_{1,r}}}$ are
determined by \Eref{g13}. Electromagnetic fields at other points
of the first layer would be computed using \eref{g12}. Inside
other layers, electromagnetic fields will be calculated using
\eref{g10} and \eref{g11}.
\section {Validation and Numerical Implementations}
In this section, several examples will be considered to illustrate
and also validate the utilization of the proposed approach.
\subsection{PCF coupler}
\begin{figure}
\begin{center}
\epsfig{file=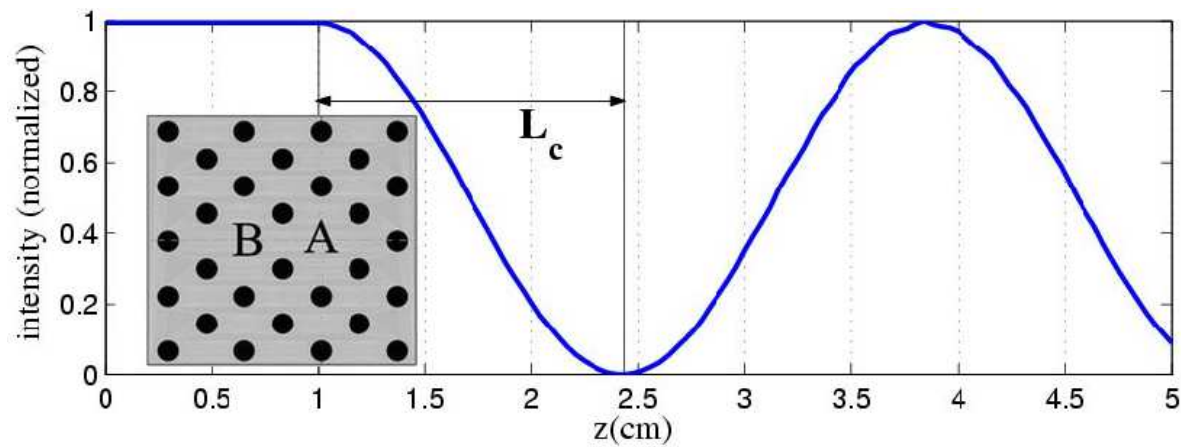,width=\columnwidth}
\end{center}
\caption{Intensity of electric field in the center of the core
$A$. The inset shows the cross section of the PCF coupler with the
two cores $A$ and $B$.} \label{fig4}
\end{figure}

The cross-section of the PCF-coupler we want to study is depicted
in the inset of \fref{fig4}. It is composed of a triangular
lattice of air-holes in silica with two missing air holes. We
validate the described approach in the present paper by verifying
the coupling action of the coupler and comparing the obtained
coupling length through this approach with the one obtained by
considering even and odd modes. In the simulation the pitch,
$\Lambda=7.2\,{\rm \mu m}$, and normalized hole-diameter to the
pitch, $d/\Lambda=0.45$ have been set. We perform the simulation
at the normalized wavelength $\lambda/\Lambda=2\pi
c/(\omega\Lambda)=0.1$.
\\ In simulation it is assumed that the light is launched into one
core of the coupler, for instance core A, by butt-coupling of a
similar single-core fibre whose core is aligned to the core A. The
coupler and the fibre coupled to it constitute a two layer medium,
which could be considered an example of the general case of
\fref{fig1}.

As it is described in Section 2.2, at first we repeat the
structure periodically in the transverse {\it xy}-plane with
$10\Lambda\times10\Lambda$ periodicity. Fiber cores of both the
single-core, first layer, and the double-core, second layer, are
considered as defects so treated by the supercell
approach~\cite{Zhi:2003}. We calculate the fundamental mode of the
single-core fiber using a MTLT-based approach
of~\cite{Dabirian:2005a} which has been briefly described in
Section 2.3. Through the simulation we obtain the fiber mode as
the voltage and current vectors ${\bf {\widehat v}_{j,inc}}(z=0)$
and $ {\bf {\widehat i}_{j,inc}}(z=0)$. These vectors describe the
electromagnetic fields of the fiber and are related to the fields
through the Eqs. (7) and \eref{g8}. Evidently from \eref{g3} the
fiber mode is the weighted summation of individual plane waves
with different wave vectors. From the known ${\bf {\widehat
v}_{j,inc}}(z=0)$ and $ {\bf {\widehat i}_{j,inc}}(z=0)$
electromagnetic fields inside all the structure will be computed
by tracking the approach described in Section 2.3.

\begin{figure}
\begin{center}
\epsfig{file=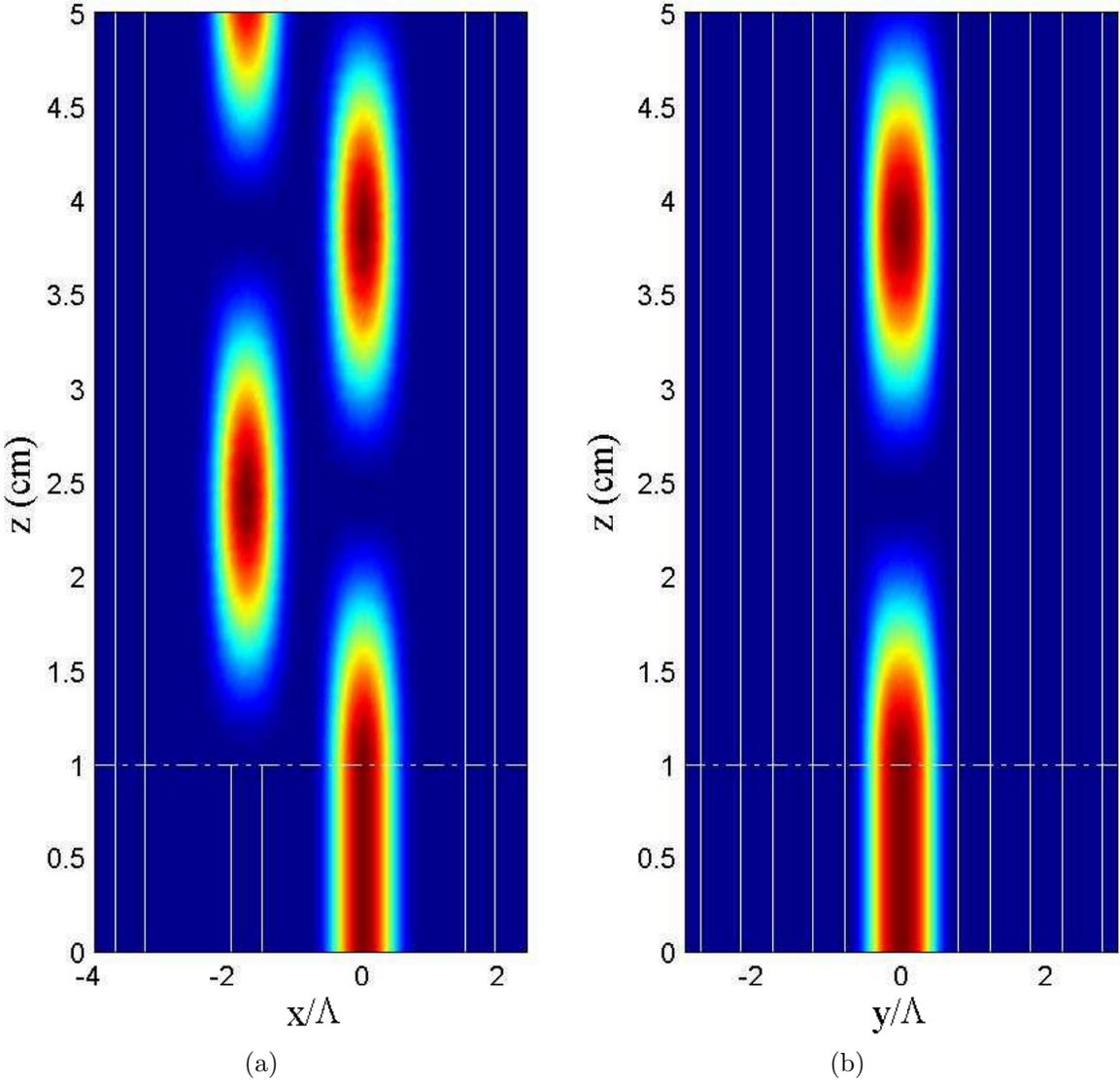,width=\columnwidth}
\end{center}
\caption{Distribution of electric field intensity (a) in the $xz$
plane at the centers of the fibre cores (b) in the $xy$ plane. The
electric field in the $xy$ plane is computed at the center of the
core $A$.} \label{fig5}
\end{figure}

We illustrate in \fref{fig5} the normalized electric field
intensity when the $HE_{11}$ mode of the single-core fibre,
travelling across the $z$-axis, is launched into the core $A$ of
the dual-core fibre. Inside the dual-core fibre, the light starts
coupling from the core $A$ to core $B$. Up to the distance of
$1440\,{\rm \mu m}$ from the interface of the coupler and
single-core fibre ($z=2.44\,{\rm cm}$) all the confined light in
the core $A$ will be coupled to the core $B$. This distance is
called coupling length, $L_c$, and alternatively may be computed
from the difference of the propagation constants of even,
$\beta_e$, and odd, $\beta_o$, modes of the dual-core fibre
through the relation of $L_c=\pi / \mid \beta_e-\beta_o \mid$. The
computed coupling length between the even and odd modes in
translational invariant system is $1410\,{\rm \mu m}$, which is in
a good agreement with that obtained through the approach of this
paper. The normalized intensity of electric field in the center of
core $A$ is depicted in~\fref{fig4}. The coupling length has been
indicated on the figure.
\subsection{PCF Bragg grating}
\begin{figure}
\begin{center}
\epsfig{file=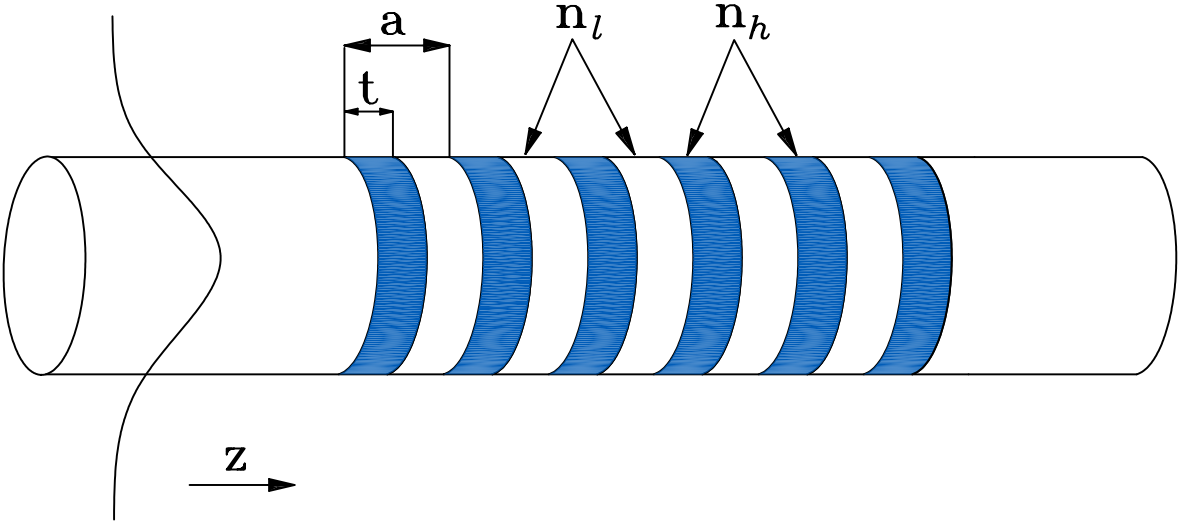,width=\columnwidth}
\end{center}
\caption{Geometry of the photonic crystal fibre distributed Bragg
reflector.} \label{fig6}
\end{figure}

\begin{figure}
\begin{center}
\epsfig{file=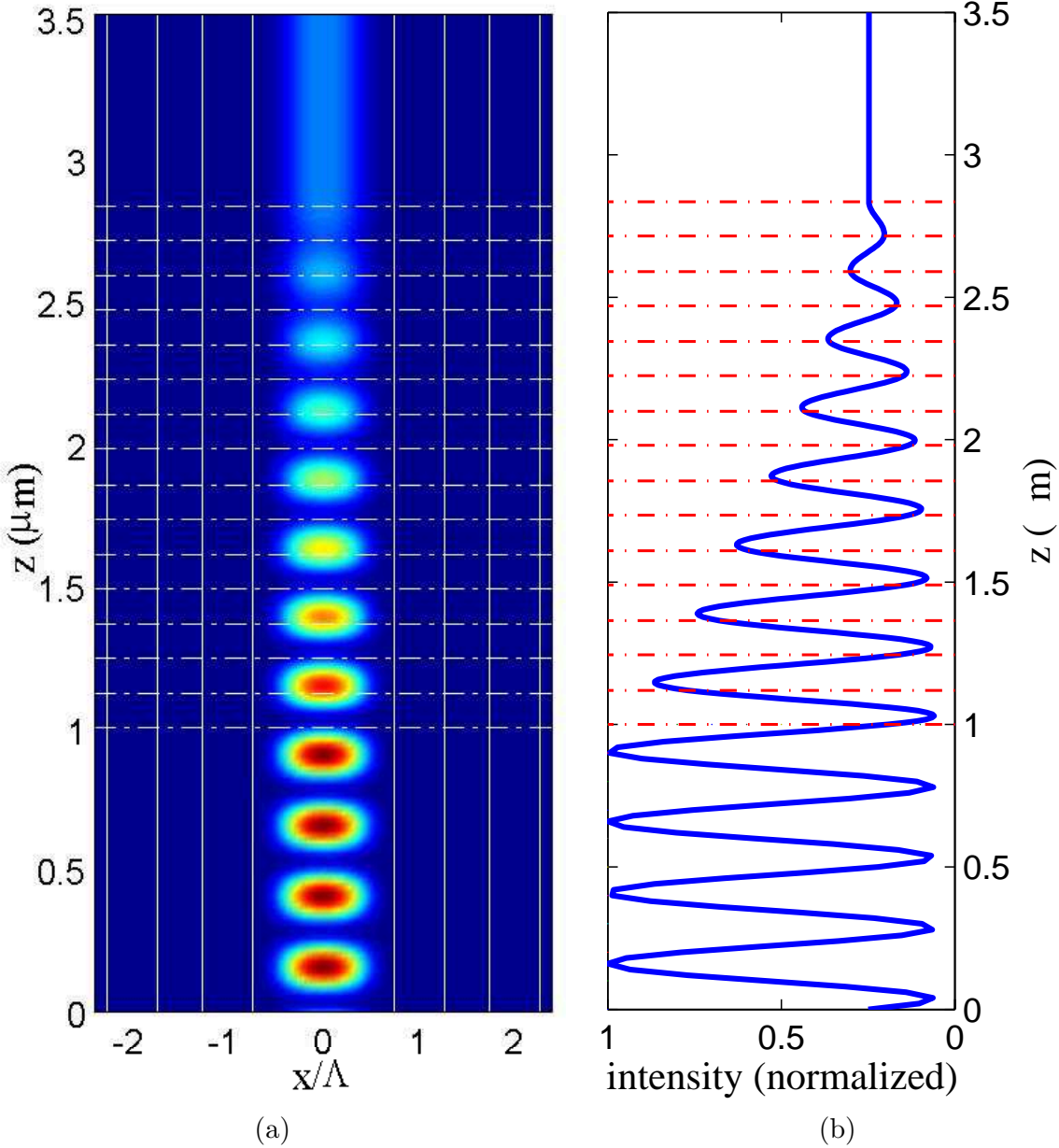,width=\columnwidth}
\end{center}
\caption {Distribution of electric field intensity (a) on the $xz$
plane at the center of the fibre core and (b) at the center of the
fibre core.} \label{fig7}
\end{figure}

The case of a PCF Bragg grating arises in various advanced
applications of photonic crystals. In PCF lasers an optical cavity
may be formed through two PCF Bragg gratings created by
introducing a spatial periodic modulation of the refractive index
to the fibre core along the fibre axis~\cite{Limpert:2003}.
Photonic crystal vertical cavity surface emitting laser
(PC-VCSEL)~\cite{Yokouchi:2003} is a novel application of the
photonic crystal in the laser application, which is similar to
standard VCSELs except that a photonic crystal structure is
defined by introducing regular lattice of air-holes with one
missing air-hole to the top mirror. These lasers, as well as the
single mode operation, have side-mode suppression ratios about
35-40dB~\cite{Song:2002}. These attractive features are
facilitated by the presence of the regular lattice of air holes as
has been studied qualitatively utilizing concepts of
PCFs~\cite{Song:2002}. Using the approach described in this paper,
the reflectivity from the top mirror could be investigated
three-dimensionally. The modelling of the laser mirrors is
generally a crucial issue in the design and analysis of
lasers~\cite{Coldren:1995}. Here we analyze an example of an
in-plane grating in a PCF to illustrate the proposed approach. The
structure under consideration is depicted in \fref{fig6}. The
cross section of each layer is a square-lattice photonic crystal
composed of air holes in the background material with one missing
air hole. The layers specified by white color have refractive
index $1.45$ and the colored ones have refractive index $1.6$. The
air-holes of normalized diameter $d/\Lambda=0.53$ are arranged on
a square lattice with pitch $\Lambda=7.2\,{\rm \mu m}$. Such
mirrors have recently been utilized as the top distributed Bragg
reflector of PC-VCSELs~\cite{Lee:2004}. In the structure the
thickness of colored layers, $t$, is $0.12\,{\rm \mu m}$ and the
periodicity of the Bragg mirror is $a=0.245\,{\rm \mu m}$.
Utilizing the approach of this paper, we examine the interaction
of the travelling fundamental mode of the first layer with the
grating at $\lambda/\Lambda=0.1$. \Fref{fig7} shows the
two-dimensional intensity plot of the electric field in the case
where the fundamental mode of the squared-lattice PCF (with
lattice index of $1.45$) is incident on the mirror. The incident
field is partially reflected at interfaces of different layers,
leading to an interference pattern caused by interference of the
incident field and the reflected ones. \Fref{fig7} also
illustrates how perfectly boundary conditions at different
material interfaces of the distributed Bragg mirror are fulfilled.
\section {Conclusions}
Optical properties of PCFs may typically be successfully analyzed
within the assumption of translational invariance along the fibre
axis. However, in real life the important device applications
employ PCFs of finite length and the hypothesis of translational
invariance is not applicable. In this work we have described a
semi-analytical approach for three-dimensional fully vectorial
analysis of photonic crystal fibre devices. Our approach rest on
the foundation of modal transmission-line theory and offers a
computationally competitive alternative to beam propagation
methods. The approach is illustrated by simulations of the
coupling action in a photonic crystal fibre coupler and the modal
reflectivity in a photonic crystal fibre distributed Bragg
reflector.
\section*{Acknowledgment}
N.~A.~M. is supported by The Danish Technical Research Council
(Grant No.~26-03-0073).

\vspace{1cm}

%\bibliographystyle{iop}
%\bibliography{PCFbibtex}

\newpage
%\listoffigures

\newpage

\newpage

\newpage

\newpage

\newpage

\newpage

\end{document}